\documentclass[12pt]{iopart}

\usepackage{graphicx}
\usepackage{bm}

\newcommand{\ket}[1]{\left\vert#1\right\rangle}
\newcommand{\s}{\uparrow}
\newcommand{\g}{\downarrow}

\begin{document}

\title{Entanglement Controlled Single-Electron Transmittivity}

\author{F Ciccarello}

\address{CNISM and Dipartimento di Fisica e
Tecnologie Relative dell'Universit\`{a} degli Studi di Palermo,
Viale delle Scienze, Edificio 18, I-90128 Palermo, Italy}
\ead{ciccarello@difter.unipa.it}

\author{G M Palma}

\address{NEST and Dipartimento di Scienze Fisiche ed Astronomiche dell'Universit\`{a} degli Studi di Palermo, Via Archirafi
36, I-90123 Palermo, Italy}

\author{M Zarcone}

\address{CNISM and Dipartimento di Fisica e
Tecnologie Relative dell'Universit\`{a} degli Studi di Palermo,
Viale delle Scienze, Edificio 18, I-90128 Palermo, Italy}

\author{Y Omar}

\address{SQIG, Instituto de Telecomunica\c{c}\~oes, P-1049-001 Lisbon and
ISEG, Universidade T\'{e}cnica de Lisboa, P-1200-781
Lisbon,Portugal}

\author{V R Vieira}

\address{CFIF and Department of Physics, Instituto Superior
T\'{e}cnico, Av. Rovisco Pais, 1049-001 Lisbon, Portugal}

\begin{abstract}
We show that the electron transmittivity of single electrons
propagating along a 1D wire in the presence of two magnetic
impurities is affected by the entanglement between the impurity
spins. For suitable values of the electron wave vector, there are
two maximally entangled spin states  which respectively make the
wire completely transparent whatever the electron spin state, or
strongly inhibits electron transmission.
\end{abstract}

\maketitle

The key role that entanglement plays not only in most quantum
information processing tasks  \cite{NC} but also in a broad range of
physical processes such as quantum transport \cite{Loss1, Buttiker,
Fazio} has been considerably clarified over the past few years. In
this letter we illustrate the interplay between entanglement and
single-electron transport properties in a 1D wire in the presence of
two scattering magnetic impurities. Such system is the electron
analogue of a Fabry-Perot (FP) interferometer \cite{rossi}, with the
two impurities playing the role of two mirrors with a quantum degree
of freedom: the spin. This suggests the interesting question whether
entanglement between the impurity spins plays any role in modulating
the transmittivity. We show that this is indeed the case. In
particular, we will show that, under simple suitable circumstances,
the maximally entangled Bell states $\ket{\Psi^+}$ and
$\ket{\Psi^-}$ of the scattering centers spins can either largely
inhibit the electron transport or make the wire completely
``transparent". Such striking behaviour, which can be intuitively
understood in terms of quantum interference, is here quantitatively
analyzed in terms of constants of motion. Furthermore, we will show
that the same scattering mechanism can be used to efficiently create
the entanglement between the impurity spins.

Let us assume to inject single conduction electrons into a clean 1D
wire - such as a semiconductor quantum wire \cite{davies} or a
single-wall carbon nanotube \cite{nanotube} - into which two
spatially separated, identical spin-1/2 magnetic impurities are
embedded (e.g. these could be realized by two quantum dots
\cite{ssqc}). Due to the presence of an exchange interaction, the
conduction electron undergoes multiple scattering with the
impurities before being finally transmitted or reflected. Let us
also assume that the electron spin state can be prepared at the
input of the wire and measured at its output (this could be achieved
through ferromagnetic contacts at the source and drain of the wire
\cite{ssqc}). To be more specific, consider a 1D wire along the
$\hat{x}$ direction with the two magnetic impurities, labeled 1 and
2, embedded at $x=0$ and $x=x_0$, respectively. Assuming that the
conduction electrons are injected one at a time (this allows us to
neglect many-body effects) and that they can occupy only the lowest
subband, the Hamiltonian can be written as
\begin{equation} \label{H}
H=\frac{p^{2}}{2m^*}+ J \, \mbox{\boldmath$\sigma$}\cdot
\mathbf{S}_{1} \,\delta(x)+ J \, \mbox{\boldmath$\sigma$} \cdot
\mathbf{S}_{2}\,\delta(x-x_{0})
\end{equation}
where $p=-i \hbar \nabla$, $m^*$ and $\bm{\sigma}$  are the electron
momentum operator, effective mass and spin-1/2 operator
respectively, $\mathbf{S}_{i}$ ($i=1,2$) is the spin-1/2 operator of
the $i$-th impurity and  $J$ is the exchange spin-spin coupling
constant between the electron and each impurity. All the spin
operators are in units of $\hbar$. Since the electron-impurity
collisions are elastic, the energy eigenvalues are simply
$E=\hbar^{2} k^2/2m^*$ ($k>0$) where $k$ is a good quantum number.
As the total spin Hilbert space is 8-dimensional and considering
left-incident electrons, it turns out that to each value of $k$
there corresponds an 8-fold degenerate energy level. Let
$\mathbf{S}=\bm{\sigma}+\mathbf{S}_{1}+\mathbf{S}_{2}$ be the total
spin of the system. Since $\mathbf{S^2}$ and $S_z$, with quantum
numbers $s$ and $m_s$, respectively, are constants of motion, $H$
can be block diagonalized, each block corresponding to an eigenspace
of fixed $s$ (for three spins 1/2, the possible values of $s$ are
$1/2, 3/2$) and $m_s=-s,...,s$. Let us rewrite Eq. (\ref{H}) in the
form
\begin{equation} \label{H2}
H=\frac{p^{2}}{2m^*}+\frac{J}{2}
\left(\mathbf{S}_{e1}^{2}-\frac{3}{2}\right)
\delta(x)+\frac{J}{2}\left(\mathbf{S}_{e2}^{2}-\frac{3}{2}\right)
\delta(x-x_{0})
\end{equation}
where $\mathbf{S}_{ei}=\bm{\sigma}+\mathbf{S}_{i}$ ($i=1,2$) is the
total spin of the electron and the $i$-th impurity. Note that, in
general $\mathbf{S}_{e1}^{2}$ and $\mathbf{S}_{e2}^{2}$ do not
commute. Here we choose as spin space basis the states $\ket{s_{e2};
s, m_s}$, common eigenstates of $\mathbf{S}_{e2}^{2}$,
$\mathbf{S}^{2}$ and $S_z$, to express, for a fixed $k$, each of the
eight stationary states of the system as an 8-dimensional column. To
calculate the transmission probability amplitude
$t_{s_{e2}}^{(s'_{e2},s)}$ that an electron prepared in the incoming
state $\ket{k}\ket{s'_{e2}; s, m_s}$ will be transmitted in a state
$\ket{k}\ket{s_{e2}; s, m_s}$ we have derived the exact stationary
states of the system. To do this, the quantum waveguide theory
approach of Ref. \cite{amjphys} for an electron scattering with a
magnetic impurity has been properly generalized to the case of two
impurities. Note that due to the form of $H$ (see Eq. \ref{H2})
coefficients $t_{s_{e2}}^{(s'_{e2},s)}$ do not depend on $m_{s}$.

Let us first consider the subspace $s=3/2$. Since in this subspace
$\mathbf{S}_{e1}^{2}$ and $\mathbf{S}_{e2}^{2}$ commute, the states
$\ket{s_{e2}; s, m_s} = \ket{1; 3/2, m_{3/2}}$  are also eigenstates
of $\mathbf{S}_{e1}^{2}$ and the effective electron-impurities
potential in Eq. (\ref{H2}) reduces to $J/4 \, \delta(x)+J/4\,
\delta(x-x_{0})$. Note that the two impurities behave as if they
were static and the scattering between electron and impurities
cannot flip the spins. The four stationary states take therefore the
simple form $\ket{\Psi_{k,1;3/2,m_{3/2}}} = \ket{\phi_{k}}\ket{1;
3/2, m_{3/2}}$, where $\ket{\phi_{k}}$ describes the electron
orbital degrees of freedom and can be easily found imposing suitable
boundary conditions at $x=0$ and $x=x_0$. This allows us to
calculate the transmission probability amplitude $t_{1}^{(1;3/2)}$
which turns out to depend on the two dimensionless parameters $kx_0$
and $\rho(E)J$, where $\rho(E)=(\sqrt{2m^{*}/E})/\pi\hbar$ is the
density of states per unit length of the wire \cite{davies}. Here,
given length limitations, we shall omit its explicit form.

Let us now consider the $s=1/2$ subspace. Here $\mathbf{S}_{e1}^2$
and $\mathbf{S}_{e2}^2$ do not commute. This is a signature of the
fact that in this space spin-flip can occur. In each of the
two-dimensional $m_{1/2}=-1/2,1/2$ subspaces, the two stationary
states are of the form
\begin{equation}  \label{ansatz}
\ket{\Psi_{k,s_{e2}';1/2,m_{1/2}}}=\sum_{s_{e2}=0,1}\ket{\varphi_{k,s_{e2}',s_{e2}}}\ket{s_{e2};
1/2, m_{1/2}}
\end{equation}
where the index $s_{e2}'=0,1$ indicates that the incident spin state
of (\ref{ansatz}) is $\ket{s_{e2}';1/2,m_{1/2}}$. In this subspace
the 8-dimensional column representing each eigenstate of $H$ has
therefore two non-vanishing components. The transmitted part of
(\ref{ansatz}) is given by $\ket{k} \left[\sum_{s_{e2}=0,1}
t^{(s_{e2}'; 1/2)}_{s_{e2}}\ket{s_{e2}; 1/2, m_{1/2}}\right]$ where
again the four coefficients $t^{( s_{e2}'; 1/2)}_{s_{e2}}$ depend on
the two parameters $kx_0$ and $\rho(E)J$. Note that since in the
$s=1/2$ subspace $\mathbf{S}_{e2}^2$ is not conserved, unlike the
$s=3/2$ case, $t^{(s_{e2}'; 1/2)}_{s_{e2}}\neq 0$ for $s_{e2}\neq
s_{e2}'$.

The knowledge of all the exact transmission amplitudes $t^{(s_{e2}';
s)}_{s_{e2}}$ allows us to determine, at all orders in the coupling
constant $J$, how an incident wave $\ket{k}\ket{\chi}$, where
$\ket{\chi}$ is an arbitrary overall spin state, is transmitted
after scattering. The state $\ket{k}\ket{\chi}$ is the incident part
of the stationary state
\begin{equation}\label{sviluppochi}
\ket{\Psi_{k,\chi}}=\sum_{s_{e2}',s,m_{s}} \langle s_{e2}';s,m_{s}
\ket{\chi} \ket{\Psi_{k,s_{e2}';s,m_{s}}}
\end{equation}
where $s_{e2}'=1$ for $s=3/2$ , while $s_{e2}'=0,1$ for $s=1/2$. The
overall electron transmittivity $T$ is obtained by expressing the
transmitted part of each stationary state
$\ket{\Psi_{k,s_{e2}';s,m_{s}}}$ in terms of $t^{(s_{e2}';
s)}_{s_{e2}}$, rearranging (\ref{sviluppochi}) as a linear expansion
in the basis $\ket{s_{e2};s,m_{s}}$ and then summing the squared
modules of the coefficients of the expansion. We are now able to
investigate how electron transmission depends on the state in which
the two impurities are prepared. Let us start with the following
family of impurity spins states
\begin{equation} \label{family2}
\ket{\Psi (\vartheta , \varphi)} = \cos\vartheta\ket{\s\g} +
e^{i\varphi}\sin\vartheta\ket{\g\s}
\end{equation}
with $\vartheta\in[0,2\pi]$ and $\varphi\in[0,\pi]$. This family
includes both maximally entangled and product states. The electron
transmittivity $T$ when the injected electron spin state is
$\ket{\s}$ with the impurities prepared in the product states
$\ket{\s\g}$ and $\ket{\g\s}$ is plotted in Figs. 1a and 1b,
respectively.
\begin{figure}[htbp]
              {\hbox{ {\includegraphics[scale=0.7]{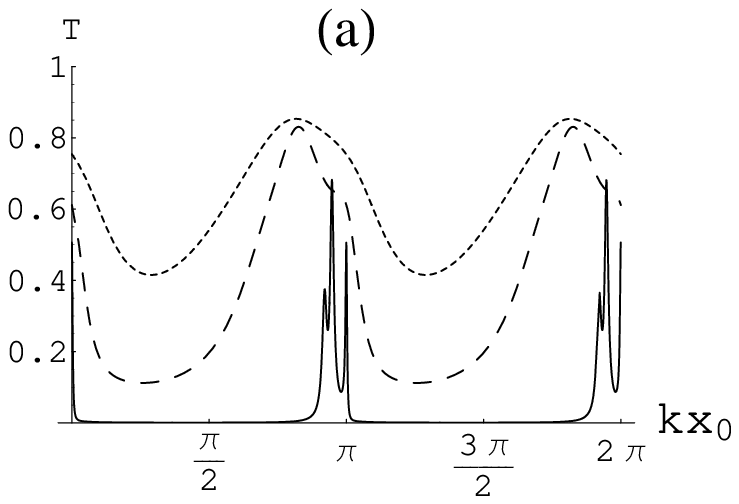}}
{\includegraphics[scale=0.7]{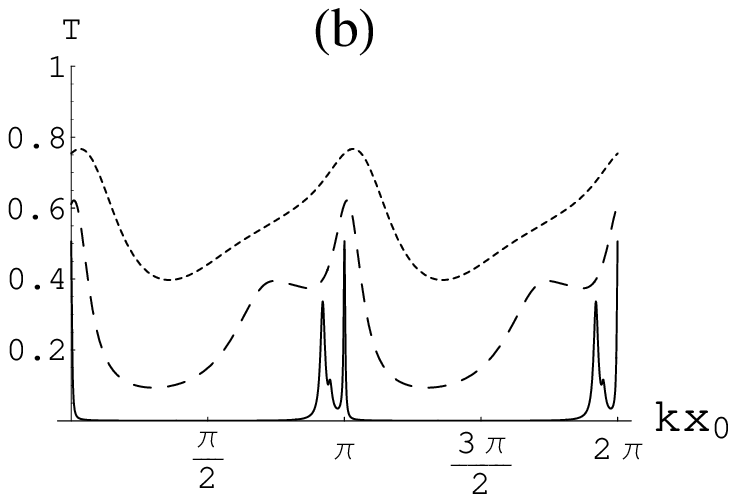}
              }}}
\caption{\footnotesize {Electron transmittivity $T$ as a function of
$kx_{0}$ when the electron is injected in the state $\ket{\s}$ with
the impurities prepared in the state $\ket{\s\g}$ (a) and
$\ket{\g\s}$ (b). Dotted, dashed and solid lines stand for
$\rho(E)J=1,2,10$, respectively.}}
\end{figure}
A behaviour similar to a FP interferometer with partially silvered
mirrors, with equally spaced maxima of transmittivity, is exhibited.
In Fig. 1a principal maxima occur around a value of $kx_{0}\neq
n\pi$ which tends to $kx_{0} =n \pi$ for increasing values of
$\rho(E)J$, while in Fig. 1b they occur at $kx_{0} =n \pi$. As
$\rho(E)J$ is increased, maxima get lower and lower and sharpen.
Remarkably, in both cases the electron and impurities spin state is
changed after the scattering and the electron undergoes a loss of
coherence, since we always have $T<1$ \cite{imry,joshi,datta}. A
similar behaviour with decoherence occurs when the two impurity
spins are prepared in the maximally entangled state $\ket{\Psi^{+}}
= (\ket{\s\g} +\ket{\g\s})/\sqrt{2}$ (see Fig. 2a). Again the
transmitted spin state differs from the incident one and, in
particular, when $kx_{0} =n \pi$, it turns out to be a linear
combination of $\ket{\s}\ket{\Psi^{+}}$ and $\ket{\g}\ket{\s\s}$.
\begin{figure}[htbp]
              {\hbox{ {\includegraphics[scale=0.7]{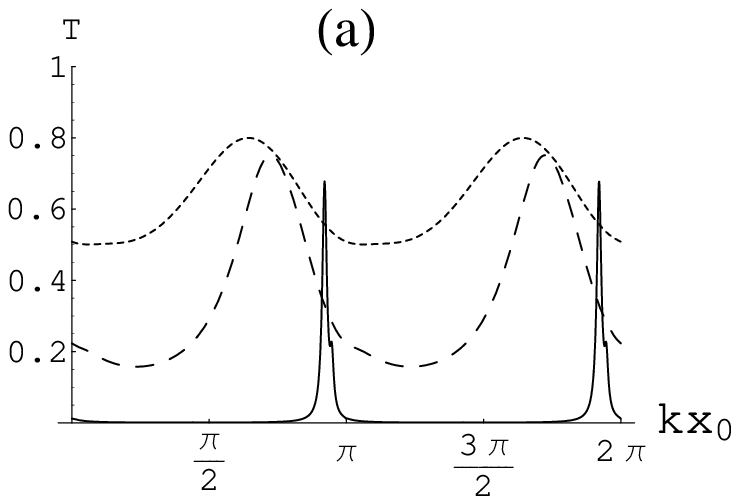}}
{\includegraphics[scale=0.7]{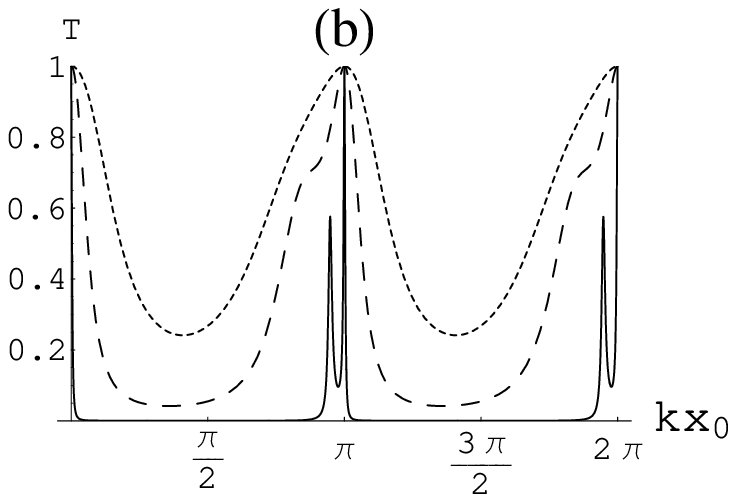}
              }}}
\caption{\footnotesize {Electron transmittivity $T$ as a function of
$kx_{0}$ when the electron is injected in the state $\ket{\s}$ with
the impurities prepared in the state $\ket{\Psi^{+}}$ (a) and
$\ket{\Psi^{-}}$ (b). Dotted, dashed and solid lines stand for
$\rho(E)J=1,2,10$, respectively.}} \label{fig3}
\end{figure}
A striking behaviour however appears when the impurity spins are
prepared in the maximally entangled state $\ket{\Psi^{-}} =
(\ket{\s\g} -\ket{\g\s})/\sqrt{2}$: as shown in Fig. 2b, the wire
becomes ``transparent" for $kx_{0}= n\pi$. In other words perfect
transmittivity $T=1$ is reached at $kx_{0}= n\pi$ regardless of the
value of $\rho(E)J$, with peaks getting narrower for increasing
values of $\rho(E)J$. Furthermore, under the resonance condition
$kx_{0}= n\pi$, the spin state $\ket{\s}\ket{\Psi^{-}}$ is
transmitted unchanged. Note that this occurs even if
$\ket{\s}\ket{\Psi^-}$ belongs to the $s=1/2$ subspace where
spin-flip is allowed. Using the explicit form of
$t_{s_{e2}}^{(s_{e2}',s)}$ it can be proved that for $kx_{0}= n\pi$
the only spin state which is transmitted unchanged is
$(\alpha\ket{\s}+\beta\ket{\g})\ket{\Psi^{-}}$, with arbitrary
complex values of $\alpha$ and $\beta$. Thus the wire becomes
transparent regardless of the electron spin state. This effect,
clearly due to constructive quantum interference, can be
quantitatively analyzed in terms of Hamiltonian symmetries. Denoting
by $\delta_{k}(x)$ and $\delta_{k}(x-x_0)$ the effective
representations of $\delta(x)$ and $\delta(x-x_0)$, respectively, in
a subspace of fixed energy $E=\hbar^{2} k^2/2m^*$, it can be easily
proved that $\delta_{k}(x)=\delta_{k}(x-x_0)$ for $kx_{0}=n\pi$.
When this occurs the non-kinetic part $V$ of $H$ in Eq. (\ref{H})
assumes the effective representation
\begin{equation}\label{Vk}
V=J \, \mbox{\boldmath$\sigma$}\cdot \mathbf{S}_{12}
\,\delta_{k}(x)=\frac{J}{2}\left(\mathbf{S}^2-\bm{\sigma}^2-\mathbf{S}_{12}^2\right)\,\delta_{k}(x)
\end{equation}
where $\mathbf{S}_{12}=\mathbf{S}_{1}+\mathbf{S}_{2}$.  This means
that, for $kx_{0}=n\pi$, $\mathbf{S}_{12}^2$ (with quantum number
$s_{12}$) becomes a constant of motion. This is physically
reasonable since the condition $kx_{0}=n\pi$ implies that the
electron is found at $x=0$ and $x=x_0$ with equal probability and,
as a consequence, the two impurities are equally coupled to the
electron spin. Furthermore, $V$ turns out to vanish for $s=1/2$ and
$s_{12}=0$. This is the case for the initial state
$(\alpha\ket{\s}+\beta\ket{\g})\ket{\Psi^{-}}$ as this is an
eigenstate of $\mathbf{S}^{2}$ and $\mathbf{S}_{12}^{2}$ with
quantum numbers $s=1/2$ and $s_{12}=0$, respectively. Therefore,
when this state is prepared and $kx_{0}=n\pi$, no spin-flip occurs
and the wire becomes transparent. This is not the case for the state
$\ket{\s}\ket{\Psi^+}$ belonging to the degenerate 2-dimensional
eigenspace of $\mathbf{S}_{12}^{2}$ and $S_{z}$ with quantum numbers
$s_{12}=1$ and $m=1/2$, respectively. As a consequence, when
$kx_{0}= n\pi$, the transmitted spin state will result in a linear
combination of $\ket{\s}\ket{\Psi^+}$ and $\ket{\g}\ket{\s \s}$,
implying spin-flip and decoherence. To further illustrate these
results we have plotted the transmittivity $T$ when the electron is
injected in an arbitrary spin state $(\alpha\ket{\s}+\beta\ket{\g})$
with the impurities prepared in a state (\ref{family2}) as a
function of $\vartheta$ and $\varphi$, for $kx_{0}= n\pi$ and
$\rho(E) J=10$ (see Fig. 3). Note how the electron transmission
depends crucially on the relative phase $\varphi$ between the
impurity spin states $\ket{\s\g}$ and $\ket{\g\s}$. The maximum
value of $T$ occurs when the impurities are prepared in the singlet
state $\ket{\Psi^{-}}$, while its minima occur for the triplet state
$\ket{\Psi^{+}}$.
\begin{figure}
 \includegraphics [scale=0.8]{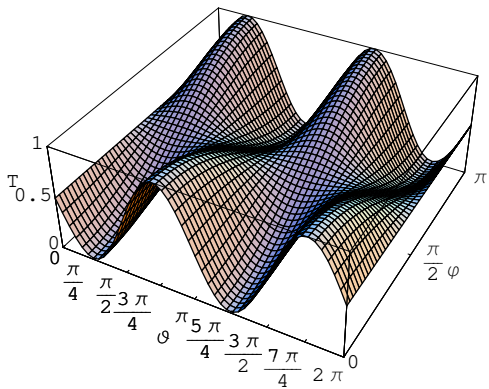}
 \caption{\label{teta_phi} Electron transmittivity $T$ at $kx_{0}=n\pi$ and $\rho(E)J=10$
 when the electron is injected in an arbitrary spin
state $(\alpha\ket{\s}+\beta\ket{\g})$ with the impurities prepared
in the state
$\cos{\vartheta}\ket{\s\g}+e^{i\varphi}\sin{\vartheta}\ket{\g\s}$.}
\label{fig4}
\end{figure}
As we have discussed, $T<1$ (it gets smaller and smaller for
increasing values of $\rho(E) J$) for $\ket{\Psi^{+}}$ due to
decoherence effects. Since the set of states (\ref{family2}) is
spanned by $\ket{\Psi^{-}}$ and $\ket{\Psi^{+}}$, the transmittivity
for a generic state (\ref{family2}) will have intermediate values
between the value of $T$ for $\ket{\Psi^{+}}$ and 1.

The most remarkable result emerging from the above discussion is
that, within the set of initial impurity spins states
(\ref{family2}), maximally entangled states $\ket{\Psi^{-}}$ and
$\ket{\Psi^{+}}$ have the relevant property to maximize or minimize
electron transmission. We have chosen $\rho(E)J=10$ to better
highlight this behaviour, but this happens for any value of
$\rho(E)J$. This result suggests the appealing possibility to use
the relative phase $\varphi$ as a control parameter to modulate the
electron transmission in a 1D wire.

For this task to be correctly performed, it is required that the
state $\ket{\Psi^+}$ in which the impurities must be prepared to
inhibit electron transmission, can be protected from spin-flip
events. This can be achieved if the electron is injected in a fixed
spin state, let us say $\ket{\s}$, and analyzed in the same state
when transmitted. Let us denote by $T_{+}$ the conditional
probability that the electron is transmitted in the state
$\ket{\s}$. In Figs. 4a and 4b we have plotted $T$ and $T_{+}$,
respectively, for an initial impurity spins state (\ref{family2})
with the electron injected in the state $\ket{\s}$ and for
$kx_0=n\pi$ and $\rho(E)J=2$. Note how the filtering can be used to
efficiently reduce the electron transmission with the impurities
prepared in the state $\ket{\Psi^+}$.
\begin{figure}[htbp]
              {\hbox{ {\includegraphics[scale=0.7]{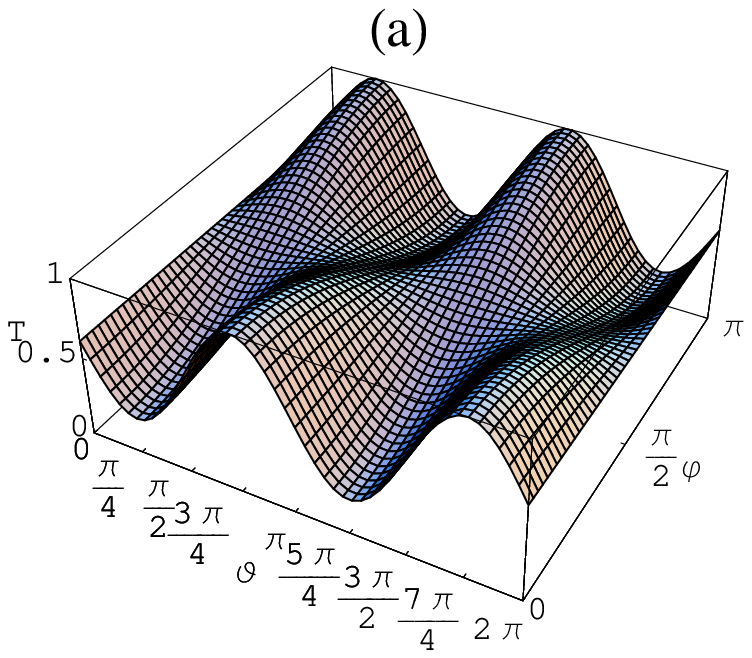}}
{\includegraphics[scale=0.7]{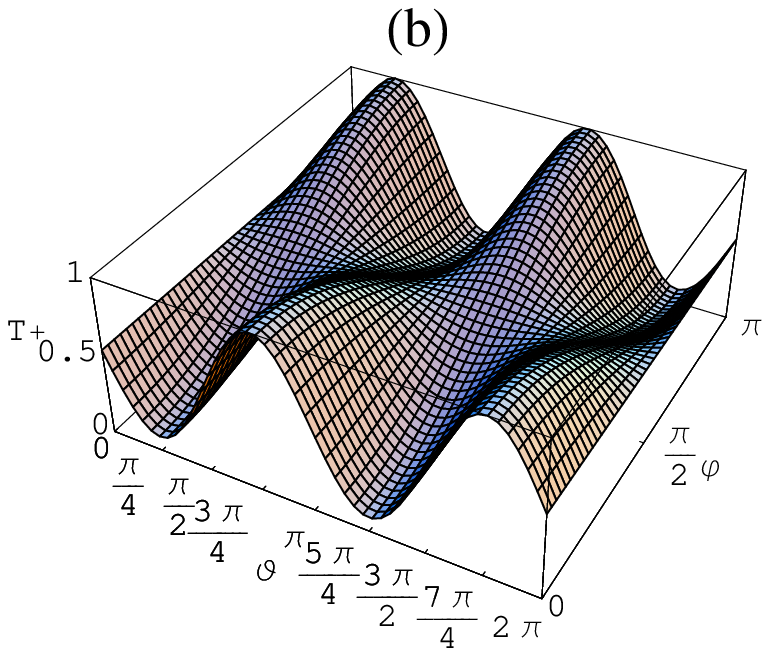}}}} \caption{\footnotesize
{Electron transmittivity $T$ (a) and conditional electron
transmittivity $T_+$ (b) at $kx_{0}=n\pi$ and $\rho(E)J=2$ when the
electron is injected in the state $\ket{\s}$ with the impurities
prepared in the state
$\cos{\vartheta}\ket{\s\g}+e^{i\varphi}\sin{\vartheta}\ket{\g\s}$.}}
\label{teta_phi_2}
\end{figure}
We found that $T_{+}\simeq T$ for high values of $\rho(E)J$ as in
Fig. 3 for $\rho(E)J=10$. Thus in these cases no spin-filtering is
required.

The above features are however not present for all sets of maximally
entangled states. Let us for instance consider the family of initial
impurity spins states
\begin{equation} \label{uu_dd}
\ket{\phi (\vartheta , \varphi)} = \cos\vartheta\ket{\s\s} +
e^{i\varphi}\sin\vartheta\ket{\g\g}
\end{equation}
with $\vartheta\in[0,2\pi]$ and $\varphi\in[0,\pi]$. Our
calculations show that, in this case, the two impurities behave as
if they were prepared in a statistical mixture of $\ket{\s\s}$ and
$\ket{\g\g}$ with weights $\cos^2 \vartheta$ and $\sin^2 \vartheta$,
respectively. This is illustrated in Figs. 5a, 5b and 5c for the
cases $\vartheta = \pi /4$, $\vartheta = 0$, $\vartheta = \pi /2$,
respectively, with arbitrary $\varphi$ and the electron injected
spin state being $\ket{\s}$. The phase $\varphi$ thus plays no role
and no interesting interference effect occurs. The reason for this
is that $\ket{\s}\ket{\s\s}$ and $\ket{\s}\ket{\g\g}$ are
eigenstates of the constant of motion $S_z$ with different quantum
numbers $m=3/2$ and $m=-1/2$, respectively and therefore, unlike the
set of states (\ref{family2}), no quantum interference effects are
possible. Finally, note that while in the cases of Figs. 5b and 5c a
loss of electron coherence is exhibited similarly to the cases of
Figs. 1a, 1b and 2a, a coherent behaviour completely analogous to a
FP interferometer with partially silvered mirrors \cite{rossi} is
observed when the impurities are prepared in the state $\ket{\s\s}$
with the electron injected in the state $\ket{\s}$ (see Fig. 5a).
Indeed, the spin state $\ket{\s}\ket{\s\s}$ belongs to the non
degenerate eigenspace $s=3/2$, $m=3/2$ where spin-flip does not
occur and the impurities behave as they were static. However, we
note that at variance with the case of Fig. 2b, $T=1$ for values of
$kx_0$ which depend on $\rho(E)J$ and only if the electron spin is
initially aligned with the spins of the impurities.
\begin{figure}
              \includegraphics [scale=1]{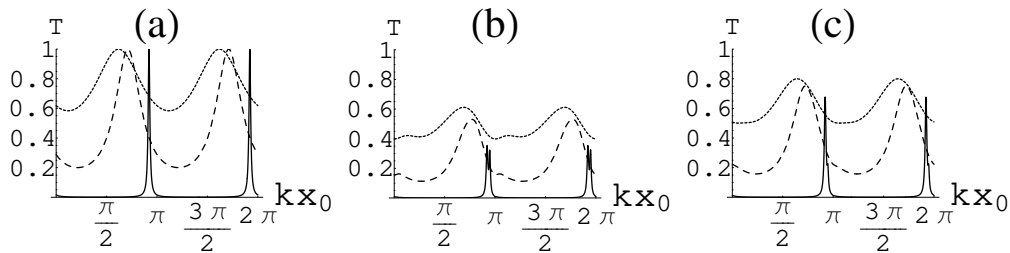}\caption{\footnotesize
{Electron transmittivity $T$ as a function of $kx_{0}$ when the the
electron is injected in the state $\ket{\s}$ with the impurities in
the initial state $\ket{\s\s}$ (a), $\ket{\g\g}$ (b) and
$(\ket{\s\s} +e^{i\varphi}\ket{\g\g})/\sqrt{2}$ for arbitrary
$\varphi$ (c). Dotted, dashed and solid lines stand for
$\rho(E)J=1,2,10$, respectively. }} \label{fig1}
\end{figure}

To conclude we show how the scattering itself can be used to
generate the desired entanglement between the impurity spins. To
modulate the electron transmission we are interested in generating
either a $\ket{\Psi^+}$ or a $\ket{\Psi^-}$ state. Such states can
be easily transformed into each other by simply introducing a
relative phase shift through a local field. A $\ket{\Psi^+}$ state
can be generated by injecting an electron in the state $\ket{\s}$
with the two impurities prepared in the state $\ket{\g\g}$, in the
spirit of \cite{yasser}. When $kx_{0}=n\pi$, due to conservation of
$\mathbf{S}_{12}^2$ and $S_z$ the transmitted spin state will be a
linear combination of $\ket{\s}\ket{\g\g}$ and
$\ket{\g}\ket{\Psi^{+}}$. An output filter selecting only
transmitted electrons in the state $\ket{\g}$ can thus be used to
project the impurities into the state $\ket{\Psi^{+}}$. The
conditional probability $T_{-}$ for this event is plotted in Fig. 6
as a function of $\rho(E)J$.
\begin{figure}
 \includegraphics [scale=0.8]{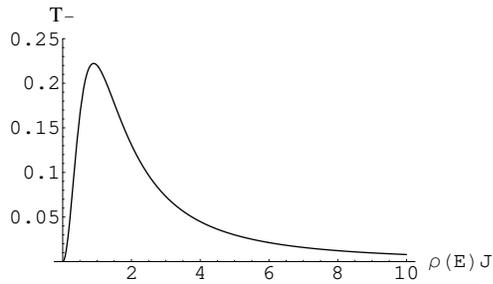}
 \caption{\label{P} Conditional electron transmittivity $T_{-}$ at $kx_{0}=n\pi$ as a function of
 $\rho(E)J$ when the electron is injected in the state $\ket{\s}$ with the impurities prepared in the state
 $\ket{\g\g}$.}
\end{figure}
A probability larger than 20\% can be reached with $\rho(E)J\simeq
1$.

To estimate the feasibility of the device here proposed, let us
assume an electron effective mass of $0.067\,m_0$ (as in GaAs
quantum wires) and the impurities to be two quantum dots each one of
size 1 nm. As a consequence, the maximum electron energy allowing to
assume a contact electron-dot potential - as in Eq. (\ref{H}) - is
around 2 meV. In this case, for $\rho(E)J\simeq 1$ we obtain
$J\simeq$ 1 eV{\AA} which appears to be a reasonable value.

Finally, we would like to point out that the above main result
showed in Fig. 3 opens the possibility of a new maximally entangled
states detection scheme. Indeed, electron transmission can be used
as a probe to detect maximally entangled singlet and triplet states
of two localized spins within the family (\ref{family2}). In
particular, it should be clear from the above discussion that in the
case of $\ket{\Psi^-}$ this would be a quantum non-demolition (QND)
detection scheme.

In summary, in this article we have considered a 1D wire into which
single electrons are injected. Such electrons undergo multiple
scattering with two spin-1/2 magnetic impurities embedded at a fixed
distance before leaving the wire. We have derived the exact
stationary states of the system thus obtaining all the necessary
transmission probability amplitudes to describe electron transport.
We have shown that for suitable electron wave vectors (independent
on the electron-impurity coupling constant) perfect transmittivity
without spin-flip takes place provided the impurity spins are
prepared in the singlet maximally entangled state. In this regime,
singlet and triplet entangled states of the localized spins are
found to maximize and minimize, respectively, electron transmission.
This suggests the appealing idea to use entanglement as a tool to
control electron transmission through a wire.

\section*{Acknowledgements}

Helpful discussions with V. K. Dugaev are gratefully acknowledged.
YO and VRV thank the support from Funda\c{c}\~{a}o para a
Ci\^{e}ncia e a Tecnologia (Portugal), namely through programs
POCTI/POCI and projects POCI/MAT/55796/2004 QuantLog and
POCTI-SFA-2-91, partially funded by FEDER (EU).

\end{document}